\documentclass[iop,apj,tighten,times]{emulateapj}
\usepackage{apjfonts} 
\usepackage{amsmath,amstext}
\usepackage{times}
\usepackage{graphicx}
\usepackage{comment}
\usepackage{latexsym,amssymb,amsbsy,amsmath}
\usepackage{amsfonts,exscale}
\usepackage{color}
\usepackage{graphicx}

\DeclareGraphicsExtensions{.pdf,.png,.jpg}

\newcommand{\ci}{C{\sc i}\phantom{.}$(J=1\rightarrow0)$}
\newcommand{\co}{CO\phantom{.}$(J=4\rightarrow3)$}
\newcommand{\coalt}{CO\phantom{.}$(J=3\rightarrow2)$}
\newcommand{\hnc}{HNC\phantom{.}$(J=5\rightarrow4)$}
\newcommand{\cn}{CN\phantom{.}$(N=4\rightarrow3)$}

\shorttitle{ALMA observations of 9io9}
\shortauthors{J.~E.~Geach et al.}

\begin{document}

\title{A magnified view of circumnuclear star formation and feedback around an AGN at $z=2.6$}

\author{J.\,E.~Geach,$\! ^1$ R.\,J.~Ivison,$^{\! 2,3}$ S.~Dye$^4$ \&
  I.~Oteo$^{3,2}$\vspace{5pt}}

\affil{$^1$School of Physics, Astronomy \& Mathematics,
  University of Hertfordshire, College Lane, Hatfield, AL10 9AB, UK. j.geach@herts.ac.uk}

\affil{$^2$European Southern Observatory,
Karl-Schwarzschild-Stra\ss{}e 2, D-85748 Garching, Germany}

\affil{$^3$Institute for Astronomy, University of Edinburgh, Royal
Observatory, Blackford Hill, Edinburgh EH9 3HJ, UK}

\affil{$^4$School of Physics and Astronomy, University of Nottingham,
University Park, Nottingham, NG7 2RD, UK}

\begin{abstract}
We present Atacama Large Millimeter/submillimeter Array 
observations of an intrinsically radio-bright ($L_{\rm 1.4\,GHz}=(1.7\pm0.1)\times10^{25}$\,W\,Hz$^{-1}$) and infrared luminous ($L_{\rm IR}\approx10^{13}L_\odot$) galaxy at
$z=2.6$. The infrared-to-radio luminosity ratio, $q=1.8$, indicates the presence of a radio loud active galactic nucleus (AGN). Gravitational lensing by two foreground galaxies at $z\approx0.2$
provides access to physical scales of approximately 360\,pc, and
we resolve a 2.5\,kpc-radius ring of
star-forming molecular gas, traced by atomic carbon \ci{} and carbon
monoxide \co{}. We also detect emission from the cyanide radical,
\cn{}. With a velocity width of 680\,km\,s$^{-1}$, this traces dense
molecular gas travelling at velocities nearly a factor of two larger
than the rotation speed of the molecular ring. While this could indicate
the presence of a dynamical and photochemical interaction between the
AGN and molecular interstellar medium on
scales of a few 100\,pc, on-going feedback is unlikely to have a
significant impact on the assembly of stellar mass in the 
molecular ring, given the $\sim$10s\,Myr depletion timescale due to star
formation.
\end{abstract}

\section{Introduction}
 
As they accrete mass, supermassive black holes (SMBHs) at the
centres of galaxies, radiating as active galactic nuclei (AGN), are
thought to regulate the growth of the surrounding stellar bulge
through energy and momentum return into the interstellar medium
(ISM, see Fabian\ 2012 for a review). Even early models recognised this could be
an important feature of galaxy evolution (Silk \& Rees\ 1998), and now AGN feedback is
an established component of the paradigm (Granato et al.\ 2004; Di Matteo et al.\ 2005; Bower et al.\ 2006; Croton et al.\ 2006; Hopkins \& Elvis\ 2010; Faucher-Gigu\'ere \& Quataert 2012; King \& Pounds\ 2015).

Observations have shown that an AGN can drive significant outflows
of gas, including the dense molecular phase, over kiloparsec scales
(Feruglio et al.\ 2010; Rupke \& Veilleux\ 2011; Sturm et al.\ 2011; Cicone et al.\ 2014, 2015;  Tombesi et al.\ 2015;  Veilleux et
al.\ 2017; Biernacki \& Teyssier 2018). Although the canonical theoretical model for the formation
and propagation of AGN-driven outflows is well understood, we still
lack a detailed empirical understanding of the astrophysics of how
an AGN actually couples to, and affects, the dense ISM on the
sub-kiloparsec scales where circumnuclear star formation occurs. The problem is exacerbated at the cosmic peak of stellar
mass and SMBH growth, $z\approx2$--$3$ (Madau \& Dickinson\ 2014), since the relevant physical
scales are generally inaccessible. Gravitational lensing of high-{\it
  z} galaxies currently provides the only route to studying this phenomenon in
the early Universe.

`9io9'\footnote{Note
   -- the name `9io9' originates from the original {\it SpaceWarps}
   image identifier. Despite being IAU non-compliant, it has a
   pleasing brevity and so we continue to use the moniker here.} was first discovered as part of the citizen science project
{\it SpaceWarps} (Marshall et al.\ 2016; More et al.\ 2016) that aimed to discover new
lensing systems in tens of thousands of deep {\it iJK$_{\rm s}$}
color-composite images covering the Sloan Digital Sky Survey Stripe
82 (Erben et al.\ 2013; Annis et al.\ 2014; Geach et al.\ 2017). Volunteers identified a spectacular red (in {\it i$-$K$_{\rm
    s}$}) partial Einstein ring ($r_{\rm E}\approx3''$) around a
luminous red galaxy at $z \approx 0.2$ (Geach et al.\ 2015). Cross-matching
with archival and existing data from 
{\it Herschel} and the Very Large Array (VLA), and through a series of follow-up observations, 9io9
was revealed to be a  submm- and radio-bright galaxy at $z = 2.553$,
with the dust emission approaching an astonishing 1\,Jy at the peak of
the spectral energy distribution. Indeed, its prominence in millimeter
maps was noted by others, who identified 9io9 as a lens candidate by
virtue of its extreme submm flux density (e.g.\ Negrello et al.\ 2010)
independently of the {\it SpaceWarps} optical/near-infrared selection (Harrington et al.\ 2016, from the combination of the {\it Planck} Catalog of Compact Sources, {\it Planck} Collaboration\ 2014, and the Herschel-Stripe 82 Survey, Viero et al.\ 2014, as well as the Atacama Cosmology Telescope, Su et al.\ 2017).

 Even taking into account the $\mu\approx10$ lensing magnification
 (Geach et al.\ 2015), 9io9 is a hyperluminous infrared ($L_{\rm
   IR}\approx10^{13}L_\odot$) and radio luminous ($L_{\rm
   1.4GHz}\approx10^{25}$\,W\,Hz$^{-1}$) galaxy. The galaxy's radio-to-infrared
 luminosity ratio, $q_{\rm IR} = 1.8$, betrays a radio-loud AGN (Ivison et al.\ 2010), but
 with copious amounts of ongoing star formation -- of the order $10^3
 M_\odot$\,yr$^{-1}$ -- contributing to $L_{\rm IR}$. With
 its remarkable brightness in the millimeter, 9io9 offers a unique
 opportunity to study the resolved properties of the cold and dense
 ISM on sub-kiloparsec scales around the central engine of a growing
 SMBH, less than 3\,Gyr after the Big Bang. 
 
 In this work we present new
 observations of 9io9 with the Atacama Large Millimeter/submillimeter Array
 (ALMA) to study the resolved properties of molecular gas in the galaxy
 across a range of densities. When calculating luminosities and physical scales we assume
 a `{\it Planck} 2015' cosmology ({\it Planck} Collaboration\ 2016).

\phantom{\ci{}}

\begin{figure*}
\centerline{\includegraphics[width=0.7\textwidth]{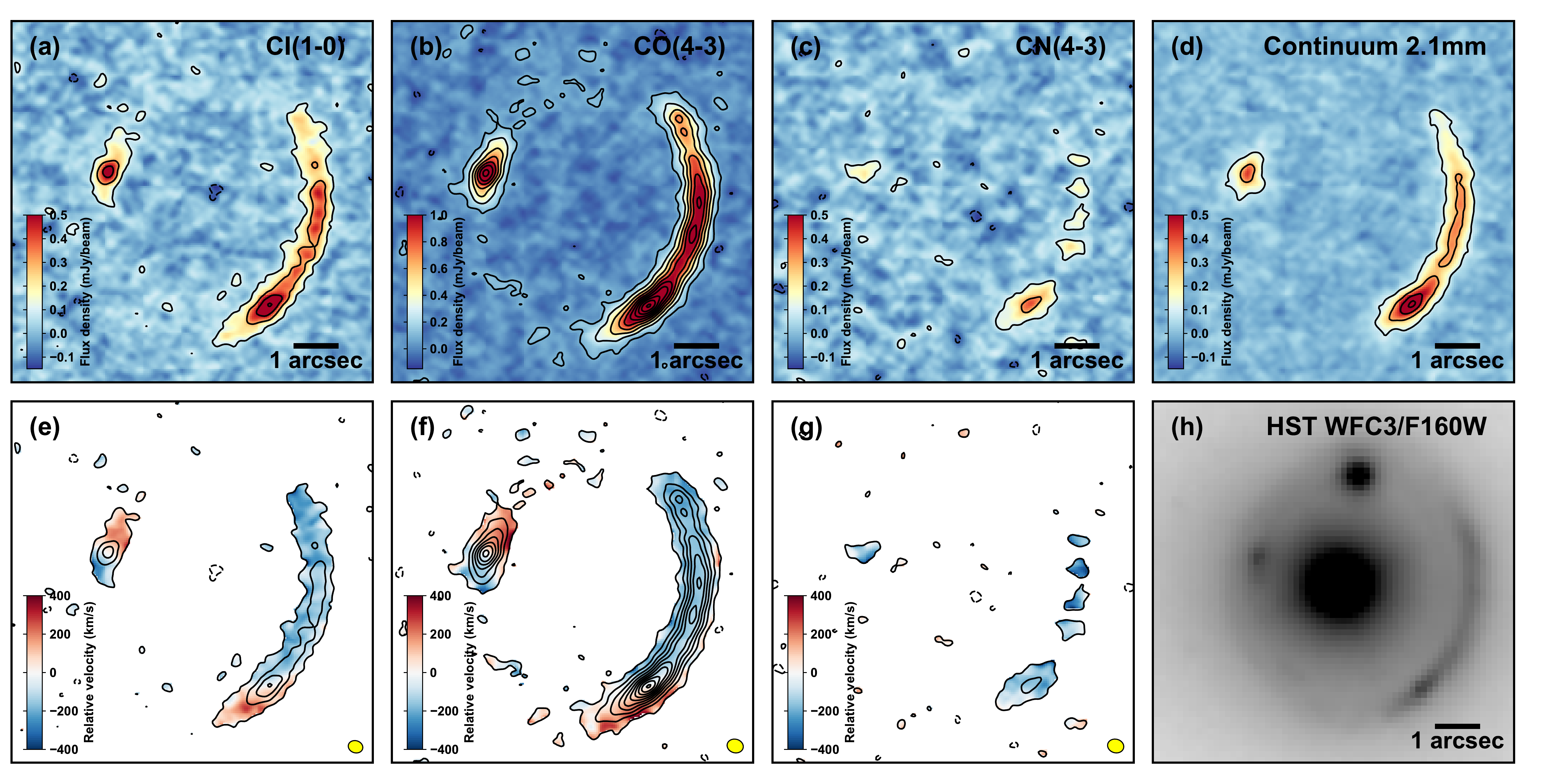}\includegraphics[width=0.265\textwidth]{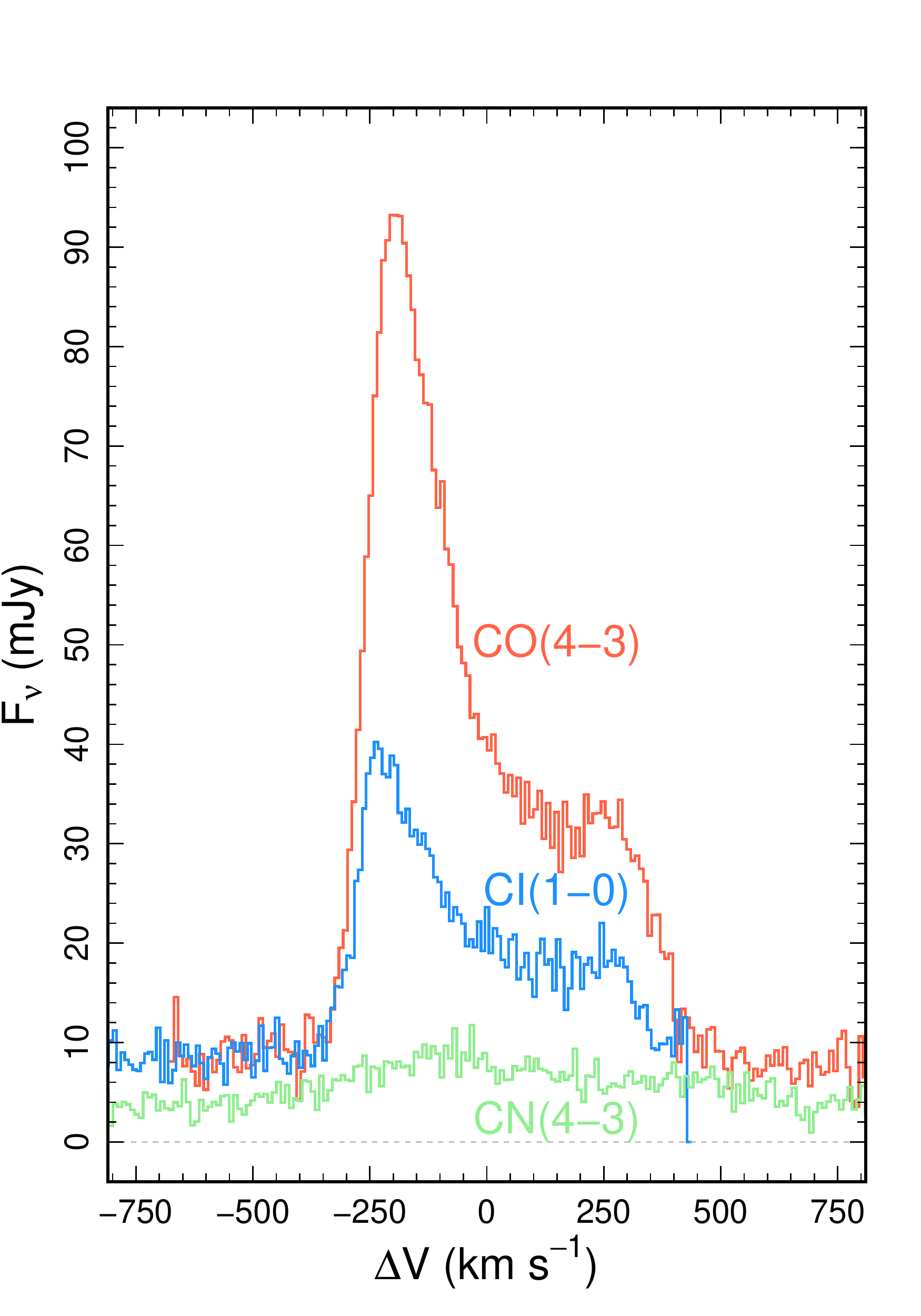}}
\caption{ALMA observations of 9io9. Panels (a--c) show the
  velocity-averaged continuum-subtracted \ci{}, \co{} and \cn{} maps
  respectively, and panel (d) shows the 2.1\,mm continuum. In panels (a--d) contours start at 100\,$\mu$Jy\,beam$^{-1}$
  and increase in steps of 200\,$\mu$Jy\,beam$^{-1}$. Panels (e--g) show the \ci{}, \co{} and \cn{}
  velocity fields derived from the 1st moment maps, where the
  blue-to-red color scale spans $-400<\Delta V
  <400$\,km\,s$^{-1}$ (black contours are the flux density contours in (a--c). The yellow ellipse indicates the size and shape
  of the synthesized beam. Panel (h) shows
  the archival {\it Hubble Space Telescope} 1600\,nm (WFC3/F160W) image of the
  target (data acquired from The Mikulski Archive for Space Telescopes, proposal ID 14653), showing the partial Einstein ring of the background galaxy
  and the two foreground lensing galaxies: a $z\approx0.2$ elliptical and
  a smaller northern companion, which has similar color and is assumed to be at the same redshift. All images are
  orientated north up, east left. The plot on the right shows the galaxy-integrated \ci{}, \co{} and \cn{} spectra, including continuum contribution. The asymmetric double horned profile of \ci{} \& \co{} are consistent with the lower spectral resolution observations of \coalt{} in 9io9 by Harrington et al.\ (2016).}
\end{figure*}

\section{Observations}

9io9 (02$^{\rm h}$09$^{\rm m}$41.3$^{s}$,
+00$^\circ$15$'$58.5$''$) was observed with the ALMA
12\,m array on 11$^{\rm th}$ and 14$^{\rm th}$ December 2017 in two
53\,minute execution blocks as part of project 2017.1.00814.S. The
representative frequency of the tuning was 139.2\,GHz in Band 4, with central frequencies of the four spectral windows 127.196, 129.071, 139.196, 141.071\,GHz. The correlator was set up in Frequency Division Mode with 480
channels per 1.875\,GHz-wide baseband with a total bandwidth of
3.6\,GHz, recording dual linear polarizations. Calibrators included
J0238$+$1636 and J0208$-$0047. Observations were designed to cover
the redshifted emission lines of the atomic carbon
$^3P_1\rightarrow\phantom{}^3P_0$ fine structure line \ci{}
($\nu_{\rm rest}=492$\,GHz) and carbon monoxide \co{} ($\nu_{\rm
  rest}=461$\,GHz), but the bandwidth also covers hydrogen isocyanide
\hnc{} ($\nu_{\rm rest}=453.270$\,GHz) and the cyanide radical,
\cn{}. The latter comprises 19 hyperfine structure components,
distributed in the three main fine structure spin groups:
$J=7/2\rightarrow7/2$, $J=7/2\rightarrow5/2$, $J=9/2\rightarrow7/2$
(at $\nu_{\rm rest}\approx\,452.628$, 453.389, and 453.606\,GHz
respectively). The relative intensities of these three groups is approximately 0.08:0.9:1, thus the $J=7/2\rightarrow7/2$ component has a negligible contribution.

After two executions that built up 1.8 hours on-source integration,
we reached a 1$\sigma$ (rms) sensitivity of 150\,$\mu$Jy per 23\,MHz
(50\,km\,s$^{-1}$) channel. The array was in configuration C43--6
with maximum baselines $\sim$3000\,m, giving a synthesized beam
$0.27''\times0.21''$ at a position angle of 88\,degrees. We make use
of the ALMA Science Pipeline produced calibrated measurement set,
and image the visibilities using the {\sc casa} (v.5.1.0-74.el7) {\it
  clean} task, employing multiscale (scales of 0$''$, 0.5$''$ and
1.25$''$) and cleaning in frequency mode. We clean down to a stopping
threshold of 1$\sigma$, and use natural weighting in the imaging.

Figure\ 1 presents the observations. We detect thermal dust continuum
emission, the \ci{} and \co{} emission lines, and a weaker broad
emission feature at $\nu_{\rm obs}\approx127.6$\,GHz which is a blend
of \hnc{} and \cn{} that we show in Section 4.2 is dominated by the
latter. The \ci{} and \co{} lines exhibit a classic double horn
profile indicative of a rotational ring or disc (e.g. Downes \& Solomon\ 1998),
and with a distinctive shear in the velocity fields,
we can spatially resolve the kinematics of the molecular gas. 

\begin{figure*}
\centerline{\includegraphics[width=0.9\textwidth]{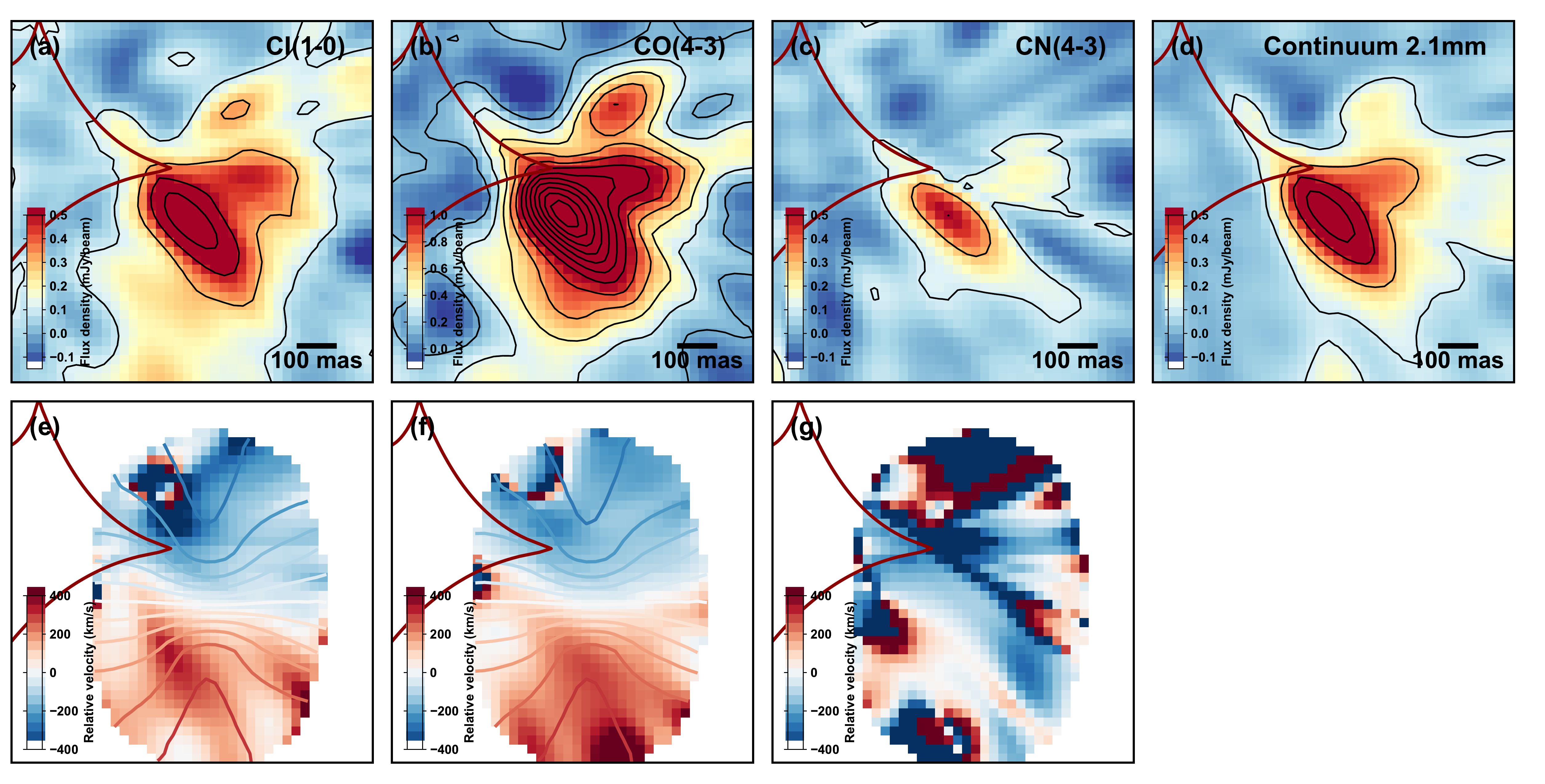}}
\caption{Source plane reconstruction of \ci{}, \co{} and \cn{} line
  emission in 9io9. Panels (a--g) show the equivalent maps as those displayed in Figure\ 1 for the image plan, with identical flux density contours. In panels (e \& f) we show
  the velocity field from our rotating ring model (Section 3.2) as
  contours, spanning $-350$--$350$\,km\,s$^{-1}$ in steps of
  $50$\,km\,s$^{-1}$. \cn{} does not exhibit a coherent velocity field
  and is largely unresolved. Red solid lines show the lensing caustic
  from our best fitting lens model (Section 3.2).}
\end{figure*}


\section{Analysis}

\subsection{Basic properties}

The sharp truncation of the \co{} and \ci{} lines characteristic of the double horn profile allow us to revise the redshift of 9io9. We find the best-fitting redshift that puts the mid-point of the full width at zero intensity ({\sc fwzi}\,$=800$\,km\,s$^{-1}$) of the continuum-subtracted lines at zero relative velocity, with $z=2.5543\pm0.0002$. This is slightly higher than the value of $z=2.553$ reported by Geach et al.\ (2015) and Harrington et al.\ (2016), but we note that the coarse velocity resolution of these previous observations might have slightly biased the redshift estimate given the asymmetric nature of the \co{} and \ci{} lines.  

We evaluate total line luminosities of a particular species, $x$ in
standard radio units as
\begin{multline}
\frac{L'_x}{{\rm K\,km\,s^{-1}\,pc^2}} = \left(\frac{3.25\times10^7}{1+z}\right)\left(\frac{D_l}{{\rm Mpc}}\right)^2\\\times\left(\frac{\nu_{\rm x,rest}}{\rm GHz}\right)^{-2}\left( \frac{\int S_\nu dV}{\rm Jy\,km\,s^{-1}} \right),
\end{multline}
\noindent where $D_l$ is the luminosity distance, $\nu_{\rm x,rest}$
is the rest frame frequency of the line and $\int S_\nu dV$ is the
velocity-integrated line flux. To evaluate $\int S_\nu dV$ for a
transition $x$ we sum over the solid angle subtended by the region
defined by the $\geq$3$\sigma$ contour in the velocity-averaged line
maps, integrated over $|\Delta V|<500$\,km\,s$^{-1}$. The uncertainty
on integrated flux (and luminosity) values is determined by adding
Gaussian noise to each channel, randomly drawn from off-line frequency
ranges of the data cube (after continuum subtraction), where the
scale is determined from the standard deviation of the flux density
in equivalent contiguous solid angle in randomly chosen source-free
parts of the data cube. By repeating this process 1000 times we
asses the standard deviation in $\int S_\nu dV$ and derived
luminosity, which we take as the 1$\sigma$ uncertainty. The source integrated flux of the \co{} and \ci{}  lines are $S\Delta V=25.9\pm0.2$\,Jy\,km\,s$^{-1}$ and $S\Delta V=9.5\pm0.1$\,Jy\,km\,s$^{-1}$, corresponding to luminosities of $\mu L'=(50.9\pm0.3)\times10^{10}$\,K\,km\,s$^{-1}$\,pc$^2$ and $\mu L'=(16.4\pm0.2)\times10^{10}$\,K\,km\,s$^{-1}$\,pc$^2$, where $\mu$ is the lensing magnification. We discuss the lens model in the next section, and the \hnc{}/\cn{} blend in Section 4.2.

\subsection{Lens modelling}

The lens model accommodates the gravitational potential of both the
primary lensing galaxy ($z\approx0.2$) and its smaller northern
companion (Geach et al.\ 2015, Figure\ 1). We use a semi-linear
inversion method (Warren \& Dye\ 2003; Dye et al. 2018) to reconstruct a pixelized
map of source surface brightness that best fits the observed lensed
image for a given lens model. The lens model is iterated,
reconstructing the source with each iteration, until the global best
fit has been obtained according to the Bayesian
evidence (Suyu et al.\ 2006). The lens mass model is motivated by the
observed lens galaxy light; for the primary lens we use an elliptical
power-law surface mass density profile of the form
\begin{equation}
\kappa=\kappa_0\,({\tilde r}/{\rm 1kpc})^{1-\alpha} \, ,
\end{equation}
where $\kappa_0$ is the normalisation surface mass density and
$\alpha$ is the power-law index of the volume mass density profile.
Here, the elliptical radius ${\tilde r}$ is defined by ${\tilde r}^2
=x^{\prime2}+y^{\prime2}/\epsilon^2$ where $\epsilon$ is the lens
elongation (i.e.\ the ratio of semi-major to semi-minor axis length)
and the co-ordinate $(x,y)$ is measured with respect to the lens
centre of mass located at $(x_c,y_c)$. The orientation of the
semi-major axis measured counter-clockwise from north is described by
the parameter $\theta$. Since the lensing effect by the secondary
galaxy on the observed image is expected to be relatively minor
(because, for realistic mass-to-light ratios, its lower observed flux
implies low mass and because the influence of the secondary mass is
largely where there is no observed Einstein ring flux) and to avoid
over-complicating the lens model, for the companion lensing galaxy we
assume a singular isothermal sphere profile fixed at the observed
galaxy light centroid with a surface mass density of the form
\begin{equation}
\kappa(r)=\kappa_{0s} \frac{r_0}{r} \, .
\end{equation}
Here, $\kappa_{0s}$ is the normalisation surface mass density and
$r_0$ is a constant set to 1\,kpc. The lens model also includes an
external shear field characterised by the shear strength, $\gamma$,
and the shear direction angle $\theta_\gamma$. 


The set of parameters is optimized using the Markov Chain Monte
Carlo (MCMC) method (Suyu et al.\ 2006) using as input the
velocity-integrated, cleaned \co{} emission. To eliminate possible
biases in the optimization, we apply the random Voronoi source plane
pixelization method (Dye et al.\ 2018). This optimal lens model is then
subsequently used to reconstruct the source plane emission on a
regular pixel grid in each observed channel to produce a source
plane data cube covering the full observed frequency range. 

We find a best fitting density profile for the primary lens that is
nearly isothermal with $\alpha=2.03$, an ellipticity of
$\epsilon=0.12$ and semi-major axis orientation of $\theta=93^\circ$
east of north which aligns closely with the observed lens galaxy
light. The model returns a total mass-to-light ratio for the
secondary lens that is 0.65$\times$ that of the primary, assuming
that both lie at the same redshift. Our lens model also includes
external shear to accommodate weaker deflections caused by the
combination of possible mass external to the primary and secondary
lens system.  The fit is improved significantly with a shear of
$\gamma=0.05$ orientated such that the direction of stretch is
$\theta_\gamma=20^\circ$ west of north. The {\sc fwhm} of the minor axis of
the effective beam in the source plane is 45\,mas, corresponding to a
physical (projected) scale of 360\,pc. Figure\ 2 shows the source
plane reconstructions of the velocity-integrated emission line maps
and velocity fields. The total magnification is $\mu=14.7\pm0.3$, and
generally the lens model is similar the one presented in Geach et
al.\ (2015). In the following, all derived physical properties are in
the $z=2.6$ source plane, and we perform the analysis in the source plane-reconstructed data cubes, thus taking into account differential lensing.

\subsection{Dynamical modelling}

We use the code {\it galpak3d} (Bouch\'e et al.\ 2015) (version 1.8.8) to
fit the source plane \co{} data cube with a rotating ring/disk
model. We make a slight modification to the publicly available code
to allow an additional definition for the density distribution,
observed in local ultraluminous infrared
galaxies (Downes \& Solomon\ 1998): 
\begin{equation}
n(R) = n_0\exp\left[4A\ln{2}\left(\frac{R-R_{\rm min}}{\Delta R}\right)^2 \right] + n_0R^\alpha,
\end{equation}

\noindent where $n_0$ and $A$ are normalisation constants, $R_{\rm
  min}$ is the inner edge of the ring (with $R_{\rm min}=0$ defining
a disk), and $R_{\rm max}$ is the outer edge (with $n(R>R_{\rm
  max})=0$). As in previous works, we fix $\alpha=0$, but $A$ is
allowed as a free parameter, as are $n_0$, $R_{\rm min}$, $R_{\rm
  max}$ and $\Delta R$. Perpendicular to the disc we assume a
Gaussian flux profile (Bouch\'e et al.\ 2015) The total velocity
dispersion of the disc is assumed to be a quadrature sum of (a) the
local isotropic velocity dispersion from self-gravity, (b) a term
due to mixing of velocities along the line-of-sight, and (c) an
intrinsic dispersion term (a free parameter) that accounts for
(e.g.) turbulent gas motions. Finally, we adopt a hyperbolic tangent
rotation curve:
\begin{equation}
V(R) = V_{\rm max}\tanh(R/R_V),
\end{equation}
\noindent where $R_V$ is the turnover radius, with $V_{\rm max}$ and
$R_V$ free parameters (Andersen \& Bershady\ 2013).  Since the effective
beam size varies over the field-of-view in the reconstructed data
cube, we convolve each channel in the input \co{} cube with a
circular Gaussian PSF with a width that aims to homogenise the
angular resolution across the source plane. We fix the {\sc fwhm} of
this kernel as the $\theta=135$\,mas which is the effective beam
size at the centre of the source plane.

We experiment with a set of several different starting parameter
values and maximum iterations to determine a `first guess' solution
that reasonably fits the data. To refine the fit, and to explore the
sensitivity of chain convergence to starting values, we take this
first set of parameters as nominal starting values, and run 1000
independent chains, each with a maximum of 5000 iterations. In each
run, we perturb each parameter by sampling from a Gaussian
distribution centred at the nominal value with a standard deviation
set at 10\% of the magnitude of the central value. We find consistent
chain convergence, with a median reduced $\chi^2=1.08$. Taking the
distribution of converged $\chi^2$ values over all 1000 chains, the
1st and 99th percentiles are $\chi^2=0.99$ and $\chi^2=1.23$
respectively. We take the 50th percentile of the converged parameters
over the 1000 chains as the final estimate of the best-fitting model
parameters, and the 16th and 84th percentiles as the 1$\sigma$
uncertainty bounds. The best-fitting model has $R_{\rm
  min}=20^{+12}_{-2}$\,mas ($161^{+102}_{-18}$\,pc), $R_{\rm
  max}=322^{+11}_{-20}$\,mas ($2647^{+88}_{-160}$\,pc),
$i=50^{+3}_{-8}$\,degrees, $\Delta R=75^{+14}_{-13}$\,mas
($613^{+111}_{-110}$\,pc), $\theta=5\pm 4$\,degrees, $V_{\rm
  max}=360^{+49}_{-11}$\,km\,s$^{-1}$, $\sigma_{\rm
  disk}=73\pm4$\,km\,s$^{-1}$. Figure\ 3 compares this model to the
data.

\begin{figure}
\centerline{\includegraphics[width=0.5\textwidth]{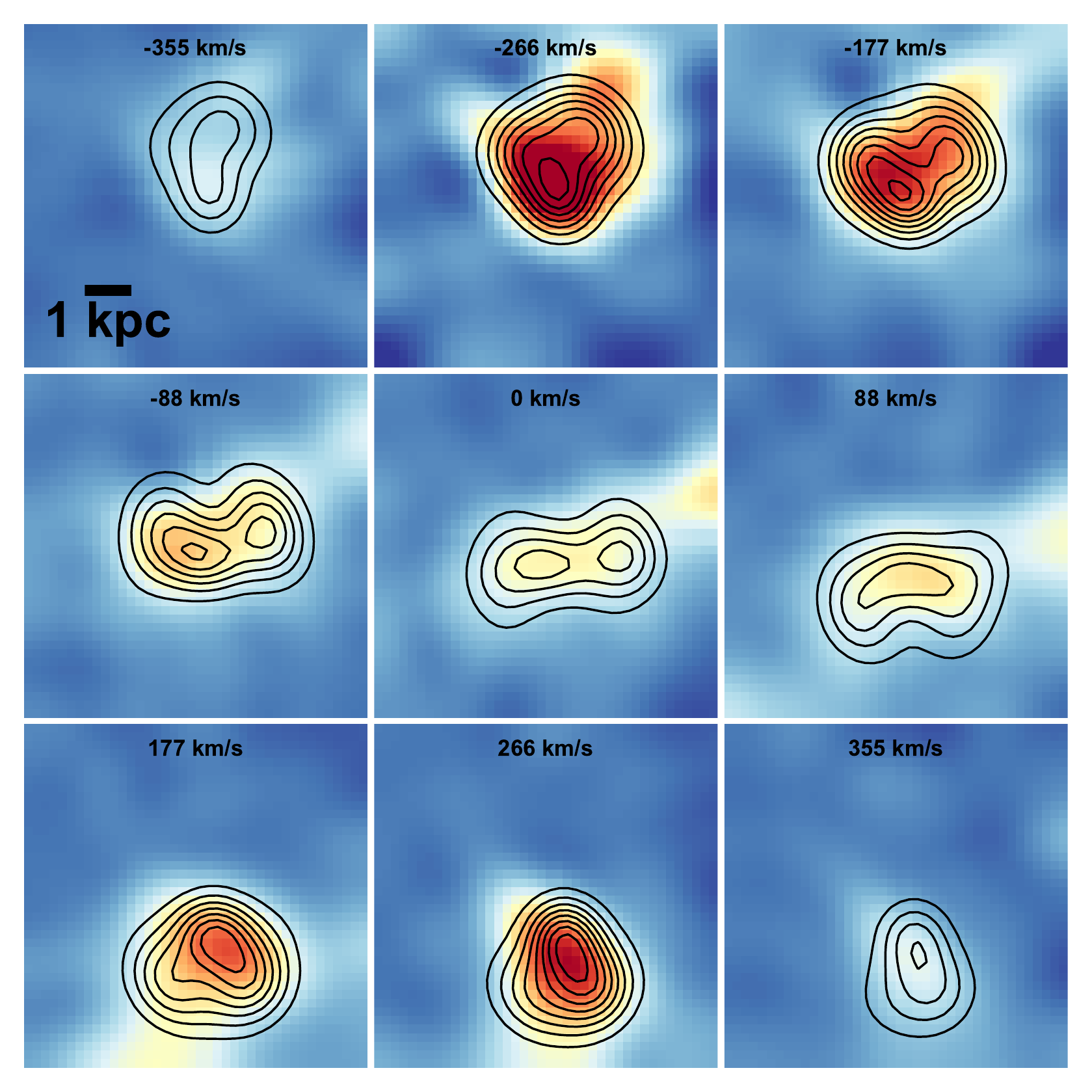}}
\caption{Dynamical modelling of 9io9. We show the
  source plane \co{} emission averaged over 88\,km\,s$^{-1}$-wide
  channels spanning $-350$--$350$\,km\,s$^{-1}$. The color scale
  saturates at 4\,mJy. Contours show the equivalent line emission
  extracted from the best-fitting model data cube, describing a
  rotating ring with inner radius 160\,pc, outer radius
  approximately 2.5\,kpc and maximum deprojected velocity
  360\,km\,s$^{-1}$. Contours show the corresponding emission in the
  model cube, starting at 0.5\,mJy and increasing in steps of
  0.5\,mJy.}
\end{figure}

\subsection{Molecular gas mass}

We measure the intrinsic (i.e.\ source plane) line luminosities as 
$L'_{\rm CI}=(8.7\pm0.1)\times10^9$\,K\,km\,s$^{-1}$\,pc$^2$ and
$L'_{\rm CO}=(32.3\pm0.2)\times10^9$\,K\,km\,s$^{-1}$\,pc$^2$. 
To evaluate the molecular hydrogen mass from the atomic carbon line
luminosity we follow previous
works (Wei\ss{} et al.\ 2003; Papadopoulos et al.\ 2004; Papadopoulos \& Greve\ 2004; Wagg et al.\ 2006; Alaghband-Zadeh et al.\ 2013; Bothwell et al.\ 2017)
\begin{multline}
\frac{M_{\rm H_2}}{M_\odot} = \left(\frac{1376}{1+z}\right)\left(\frac{D_l}{\rm
  Mpc}\right)^2\left(\frac{X_{\rm
    CI}}{10^{-5}}\right)^{-1}\\ \times\left(\frac{A_{10}}{10^{-7}{\rm
    s^{-1}}}\right)^{-1}Q^{-1}_{10}\left(\frac{S_{\rm CI}}{{\rm
    Jy\,km\,s^{-1}}}\right),
\end{multline}
\noindent where $X_{\rm CI}$ is the atomic carbon to molecular
hydrogen abundance ratio, $A_{10}$ is the Einstein $A$-coefficient
for the \ci{} transition ($A_{10}=7.93\times10^{-8}$\,s$^{-1}$) and
$Q_{10}$ is the excitation factor defined by the ratio of the column
density of the upper excited level $(^3P_1)$ to the ground state
($^3P_0$). This depends on the density and kinetic temperature of the
gas, which are not well constrained, but as in other works we assume a
C{\sc i}/H$_2$ abundance $X_{\rm CI}=3\times10^{-5}$ and excitation
$Q_{10}=0.5$ (Papadopoulos \& Greve\ 2004). This gives $M_{\rm
  H_2}=(7.5\pm 0.1)\times10^{10}M_\odot$. Recently, Rivera et al.\ (2018) presented an analysis of the \coalt{} map of 9io9 at approximately 1$''$ resolution, deriving a slightly lower $M_{\rm
  H_2}=(3.5\pm 0.2)\times10^{10}M_\odot$, however systematic uncertainties on $\alpha_{\rm CO}$ and lensing magnification could easily bring the values into parity.

\section{Interpretation}

\subsection{The molecular ring}

The total mass within $R_{\rm max}$ can be estimated from the
dynamical mass, $M_{\rm dyn}=V_{\rm max}^2R_{\rm max}/G$, with the
model giving $M_{\rm dyn}=(8.1^{+1.5}_{-0.8})\times10^{10}M_\odot$.
This indicates that the potential within 2.5\,kpc of the SMBH
is molecular gas-dominated. The line ratio $L'_{\rm CO(4-3)}/L'_{\rm CI}$ can
further inform us about the conditions of the gas in the ring, since
it is sensitive to the the dense-to-total molecular gas ratio
$f_{d/t}$, in turn thought to be a reliable indicator of the star
formation efficiency (Papadopoulos \& Geach\ 2012). For example, $L'_{\rm
  CO(4-3)}/L'_{\rm CI}$ can vary by an order of magnitude between
different star-forming environments, with values of $\sim$0.5 for
quiescent disks and clouds in the Milky Way and local Universe, up to $\sim$5 for galactic nuclei, ultraluminous infrared
galaxies and quasars (e.g.\ Israel et al. 1995, 1998;  Barvainis et al.\ 1997; Petitpas \& Wilson 1998; Israel \& Baas 2001, 2003;  Papadopoulos \& Greve 2004; Papadopoulos et al. 2004).

We measure $L'_{\rm CO(4-3)}/L'_{\rm CI}=3.7\pm 0.1$, indicating
$f_{d/t}\sim0.5$ (Papadopoulos \& Geach\ 2012). From this we can estimate
the total star formation rate by assuming the total molecular gas
mass is related to on-going star formation as $\psi_{\star} =
\epsilon f_{d/t}M_{\rm H_2}$. The factor $\epsilon$ describes the star
formation efficiency of dense molecular gas, with compelling
observational evidence that $\epsilon$ is roughly constant, possibly
reflecting a local efficiency at the fundamental scale of star
formation (Thompson et al.\ 2005). Expressed in terms of the emergent
infrared luminosity, $\epsilon=500\,(L_\odot/M_\odot)$
(Shirley et al.\ 2003; Scoville et al.\ 2004), we estimate the total star
formation rate in the molecular ring as
$\psi_\star\approx2800$\,$M_\odot$\,yr$^{-1}$, modulo systematic
uncertainties on the form of the stellar initial mass function,
(e.g.\ Zhang et al.\ 2018).

\subsection{A possible dense molecular outflow}

The broad emission feature at $\nu_{\rm obs}=127.6$\,GHz is a blend of
\hnc{} and \cn{}. The velocity-integrated emission is more compact
than the \co{} and \ci{} in the source plane (Figure\ 2), with
approximately 80\% of the integrated flux unresolved, corresponding
to emission on scales below 360\,pc. To model this feature we assume
that both \hnc{} and \cn{} contribute to the observed emission line,
and, since they trace similar gas densities
($n\sim10^{5-6}$\,cm$^{-3}$), assume that both lines are kinematically
broadened by the same Gaussian $\sigma_{\rm dense}$.

For the \cn{} line, with its various fine- and hyperfine structure
components, we assume local thermodynamic equilibrium and optically
thin emission, adopting the relative line intensities from the
Cologne Database for Molecular Spectroscopy (Muller et al.\ 2005),
calculated at 300\,K (since we have no reliable estimate of the
temperature). We fit the spectrum allowing the \hnc{} and \cn{}
amplitudes, $\sigma_{\rm dense}$ and redshift (assuming the same for
both species) to vary as free parameters. We also allow for a
constant amplitude continuum. The best fitting redshift is $z=2.5543\pm0.0005$, consistent with the value reported in Section 3.1. 

The data and best fitting model are
shown in Figure\,4, where we find that the observed emission is
dominated by \cn{} with a statistically insignificant \hnc{}
contribution. Interestingly, the opposite was found for
APM\,08279$+$5255 ($z=4$) by Gu\'elin et al.\ (2007), who found a line
blend dominated by \hnc{}, with $L'_{\rm HNC}/L'_{\rm CN}\approx2$,
although \cn{} is only tentatively detected in that system. However, we caution the reader that the deblending is highly dependent on redshift. For example, fixing $z=2.5534$ (e.g.\ Harrington et al.\ 2016) results in a more substantial \hnc{} contribution to the blend. Since internal motions/offsets of order 100\,km\,s$^{-1}$ (relative to the molecular ring) for the dense gas traced by \hnc{} and \cn{} are possible, we are reticent in drawing any conclusions regarding the relative strengths of these lines; deeper observations will be required to properly model the complex, ideally with coverage of other dense gas tracers to properly constrain the physical conditions. Regardless of this, it is clear that one, or both, of the lines must be very broad, and this might provide clues as to the nature of the very dense gas in 9io9 compared to the molecular ring.

\begin{figure}
\centerline{\includegraphics[width=0.6\textwidth]{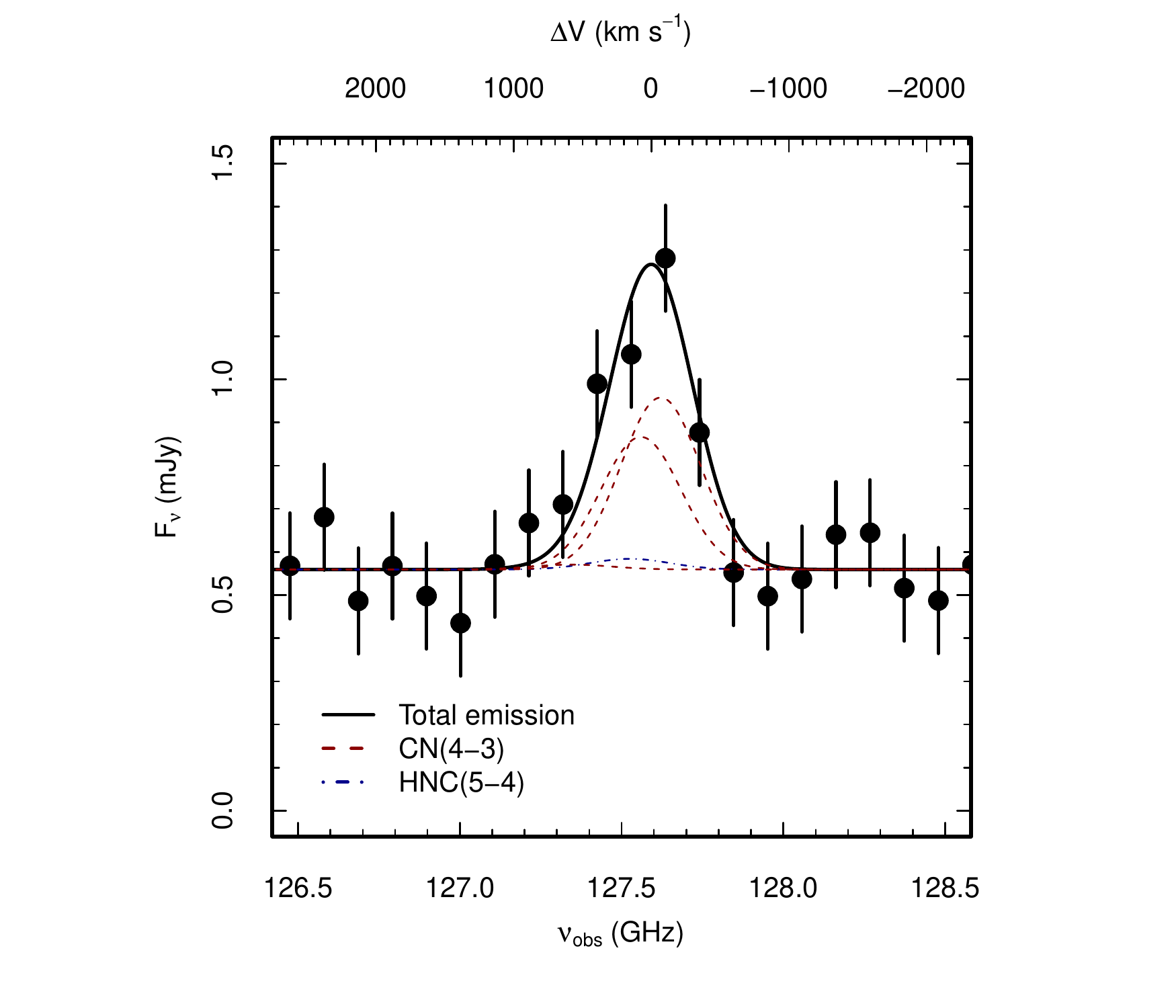}}
\caption{Detection of \cn{}. The observed spectrum at
  126.5--128.5\,GHz is modelled with a combination of \hnc{} and
  \cn{} gaussian emission lines with fixed redshift and velocity
  dispersion, plus a constant continuum component. \cn{} comprises
  19 hyperfine components in three spin groups, with the
  $J=7/2\rightarrow5/2$ and $J=9/2\rightarrow7/2$ components
  dominating the \cn{} total emission for relative intensities
  calculated assuming optically thin conditions and local
  thermodynamic equilibrium (Muller et al.\ 2005). Our best-fitting
  model shows a negligible contribution from \hnc{} (although we caution against drawing any conclusions regarding the relative strength of the lines, see Section 4.2), and supports a
  broad velocity width of {\sc fwhm~}$= 680$\,km\,s$^{-1}$. Note, for
  presentation, the data has been binned to a channel width of
  100\,MHz or 240\,km\,s$^{-1}$, but the fit was done on the full
  resolution spectrum.}
\end{figure}

The model line width is $\sigma_{\rm dense}=(289\pm48)$\,km\,s$^{-1}$,
corresponding to a {\sc fwhm~}$\approx 680$\,km\,s$^{-1}$, nearly a factor of two larger than the
deprojected maximum rotation speed of the molecular ring. Note that
fixing the velocity width of the \hnc{} and \cn{} components to the
$\sigma_{\rm disk}=73$\,km\,s$^{-1}$ dispersion of the molecular ring
does not result in a sensible fit to the data. The compact nature of
the \cn{} emission compared to \ci{} and \co{}, coupled with its large
velocity width compared to the deprojected rotation speed of the ring,
suggests that the gas traced by \cn{} does not trace the bulk of the
gas reservoir and could be dynamically decoupled from the ring.

\section{Conclusion}

One interpretation of these observations is that the gas traced by
\cn{} is outflowing, potentially indicating interaction between the
AGN and the inner part of the molecular ring or smaller scale
circumnuclear disk. However, we note that the high star formation rate
density of the ring could also be conducive to the formation and
excitation of CN (although in that case the broad line would likely
have to be produced by a supernova-driven wind; not implausible given
the high $\psi_\star$). With the current data (i.e.\ the lack of a
broader range of tracers) we cannot unambiguously distinguish between
an AGN versus `star formation' origin of the \cn{} emission; indeed
the interpretation of this species in general is rather complex
(Meijerink \& Spaans\ 2005, Wilson\ 2018). Nevertheless, CN can be produced through
the photodissociation of species such as HCN and its isomers in
environments with intense ultraviolet or X-ray radiation
fields (Fuente et al.\ 1993; Rodriguez-Franco et al.\ 1998; Meijerink \& Spaans 2005; Meijerink et al.\ 2007) as exists
around AGN (Aalto et al.\ 2002; Chung et al.\ 2011).

Given the prevalence of molecular outflows on similar scales in local
ULIRGs (e.g.\ Feruglio et al.\ 2010; Sturm et al.\ 2011; Cicone et al.\ 2014), the presence of a dense molecular outflow in 9io9 is not
surprising. Perhaps more surprising is the realisation that AGN
feedback will do little to curtail the ongoing rapid stellar mass
assembly in the surrounding ring, given the short gas consumption
timescale due to star formation, $M_{\rm
  H_2}/{\psi}_\star\sim10$s\,Myr. We cannot yet estimate the mass
outflow rate in the putative molecular wind, but it is clear that it
cannot represent a significant fraction of the total gas
reservoir. Thus, gas exhaustion, rather than quenching, will result in
9io9 transitioning into a passive elliptical galaxy. This is not to
say that the AGN will not play a regulatory role in future stellar
mass growth, but these observations suggest that co-eval radio-mode
AGN feedback could be extraneous to the rapid assembly of stellar
bulges at the peak epoch of galaxy formation.

\section*{Acknowledgements}
We thank the anonymous referee for a constructive report. J.E.G. is supported by a Royal Society University Research
Fellowship. R.J.I. and I.O. acknowledge support from the European
Research Council in the form of Advanced Investigator Programme,
COSMICISM, 321302. S.D. is supported by the UK Science and Technology
Facilities Council Ernest Rutherford Fellowship scheme. The authors
thank Susanne Aalato, Jim Dale, Jan Forbrich, Thomas Greve, Martin
Hardcastle, Mark Krumholz, Padelis Papadopoulos, Dominik Riechers and
Serena Viti for helpful discussions, and to Nicolas Bouch\'e for
advice on the use of the {\it galpak3d} code. 

This paper makes use
of the following ALMA data: ADS/JAO.ALMA\#2017.1.00814.S. ALMA is a
partnership of ESO (representing its member states), NSF (USA) and
NINS (Japan), together with NRC (Canada), MOST and ASIAA (Taiwan),
and KASI (Republic of Korea), in cooperation with the Republic of
Chile. The Joint ALMA Observatory is operated by ESO, AUI/NRAO and
NAOJ. Some of the data presented in this paper were obtained from the Mikulski Archive for Space Telescopes (MAST). STScI is operated by the Association of Universities for Research in Astronomy, Inc., under NASA contract NAS5-26555. This research has made use of the University of Hertfordshire
high-performance computing facility (http://stri-cluster.herts.ac.uk).

\end{document}